\documentclass[reprint,amsmath,amssymb,aps,pra]{revtex4-1}
\usepackage{graphicx}% Include figure files
\usepackage{dcolumn}% Align table columns on decimal point
\usepackage{bm}% bold math
\usepackage{mathrsfs}
\usepackage{amsmath} % split in equation
\usepackage{color,xcolor}
\usepackage[colorlinks=true, linkcolor=blue,urlcolor=blue,anchorcolor=blue,citecolor=blue,bookmarksnumbered]{hyperref}

\begin{document}

\title{{Enhancing tripartite photon-phonon-magnon entanglement by synergizing parametric amplifications}}

\author{Yan Wang$^{1,2}$}
\author{Jin-Lei Wu$^{3}$}
\author{Ya-Feng Jiao$^{1,2}$}
\author{Tian-Xiang Lu$^{5}$}
\author{Hui-Lai Zhang$^{1,2}$}
\author{Li-Ying Jiang$^{1,2}$}
\author{Le-Man Kuang$^{2,4}$}
\author{Hui Jing$^{2,4}$}
\email{E-mail:jinghui73@foxmail.com}
\affiliation{$^1$ School of Electronics and Information, Zhengzhou University of Light Industry, Zhengzhou, 450001, China\\
$^2$ Academy for Quantum Science and Technology, Zhengzhou University of Light Industry, Zhengzhou, 450001, China\\
$^3$ School of Physics and Microelectronics, Zhengzhou University, Zhengzhou, 450001, China\\
$^4$ Key Laboratory of Low-Dimensional Quantum Structures and Quantum Control of Ministry of Education, Department of Physics and Synergetic Innovation Center for Quantum Effects and Applications, Hunan Normal University, Changsha, 410081, China\\
$^5$ College of Physics and Electronic Information, Gannan Normal University, Ganzhou, 341000, China
}

\date{\today}

\begin{abstract}
Tripartite entanglement as a remarkable resource in quantum information science has been extensively investigated in hybrid quantum systems, whereas it is generally weak and prone to be suppressed by noise, restricting its practical application in quantum technologies. Here, we propose how to enhance the tripartite entanglement among magnons, photons and phonons in a hybrid cavity-magnon optomechanical system by exploiting a synergistic effect of the optical parametric amplification (OPA) and mechanical parametric amplification (MPA). We find that in the case of individually applying the OPA or MPA, the tripartite entanglement can be enhanced by several folds. Remarkably, upon suitably tuning the phase matching of the two parametric fields presented simultaneously, the strengths of the entanglement can be further enhanced due to the constructive interference between the OPA and MPA. We also show that our method can improve the robustness of the entanglement against thermal noise. Our work provides a promising method for manipulating the entanglement with easy tunability and may serve as a useful tool for the enhancement and protection of fragile quantum resources.

\end{abstract}
\maketitle
\section{Introduction}
Quantum entanglement \cite{RevModPhys.81.865} is arguably an essential resource in quantum technologies including quantum metrology\cite{RevModPhys.90.035005,PhysRevLett.114.110506}, quantum computation\cite{graham2022multi,bartolucci2023fusion} and quantum communication \cite{ursin2007entanglement,piveteau2022entanglement}, and it plays a central role in understanding the classical-to-quantum boundary \cite{10.1063/1.882326}. The rapid progress in experimental techniques enables direct observation of entanglement in various experimental platforms, extended from microscopic quantum units, such as multiples of single ions \cite{PhysRevLett.130.050803,hrmo2023native}, atoms \cite{van2022entangling,bluvstein2022quantum,PhysRevLett.129.050503} and photons \cite{ge2024quantum,dai2022topologically}, to massive macroscopic systems \cite{kotler2021direct,mercier2021quantum}. In addition to the efforts towards generating and utilizing the fundamental bipartite entanglement, multipartite entanglement has recently attracted increasing research interests for the perspective of exploring future quantum technologies, especially the quantum internet and programmable quantum networks \cite{armstrong2012programmable,mccutcheon2016experimental,cai2017multimode}. Recently, considerable advances have been made in preparing and manipulating tripartite entanglement in hybrid quantum systems \cite{Kurizki3866}. Examples include magnon-photon-phonon entanglement in hybrid cavity magnomechanical systems \cite{PhysRevLett.121.203601,PhysRevB.108.024105,Zuo_2024}, atom-light-mirror and light-mirror-mirror entanglement \cite{PhysRevA.77.050307,PhysRevLett.129.063602,jiao2023tripartite} in hybrid optomechanical systems. Nonetheless, limited by weak interactions, preparing highly entangled states in hybrid quantum systems is still challenging due to the detrimental effects of noise. This restricts practical application of entanglement as a resource for a wide variety of quantum technologies.

In order to protect the fragile entangled states, various schemes have been developed, e.g., by exploiting synthetic gauge fields \cite{PhysRevLett.129.063602,liu2023phase}, reservoir engineering \cite{PhysRevLett.110.253601}, dark-mode or
feedback control \cite{PhysRevResearch.4.033112,PhysRevA.106.063506,PhysRevA.95.043819}, chiral light-matter interaction \cite{patrick2024chirality}, and optical nonreciprocity \cite{PhysRevLett.125.143605,PhysRevApplied.18.064008,PhysRevA.110.012423}, etc. Over the past few years, parametric amplification has emerged as a powerful tool for quantum applications \cite{qin2024quantum}, where it has been used to significantly enhance and controllably manipulate the interactions between quantum objects in various coupled systems ranging from optomechanics \cite{PhysRevLett.114.093602,lemonde2016enhanced,PhysRevA.100.062501,zhao2020weak,wang2020enhanced,lu2024quantum}, cavity or circuit QED \cite{PhysRevLett.120.093601,PhysRevLett.120.093602,PhysRevA.99.023833,PhysRevLett.124.073602,PhysRevLett.125.203601,PhysRevA.101.053826,PhysRevA.102.032601,PhysRevLett.127.093602,PhysRevLett.126.023602,PhysRevLett.128.083604,villiers2023dynamically}, trapped ions \cite{PhysRevLett.122.030501,burd2021quantum,PhysRevA.107.032425} to hybrid spin-mechanical and magnon-mechanical systems \cite{PhysRevLett.125.153602,PhysRevApplied.17.024009,PhysRevA.107.023722,PhysRevLett.130.073602,wang2023quantum}. Particularly, optical parametric amplification (OPA) has recently been proved to be effective for improving optomechanical bipartite entanglement \cite{PhysRevA.100.043824,PhysRevA.101.033810}, and light-mirror-mirror \cite{jiao2023tripartite} and magnon-photon-phonon tripartite entanglement \cite{PhysRevA.105.063704}, which opens up a new prospect for the manipulation and enhancement of entanglement. This idea was further extended to combining the OPA and the mechanical parametric amplification (MPA) to explore the dynamics of entanglement in a cavity optomechanical system, based on which a novel method for achieving further enhancement of optomechanical entanglement was proposed very recently \cite{yang2024multi}. Given the unique importance of multipartite entanglement and the fact that the improvement of its performance is still insufficient in previous studies, it is natural to explore more efficient means to enhance and manipulate multipartite entanglement as well as improve its robustness to noise.

In this work, we propose a synergistic strategy that exploits the interplay of OPA and MPA to manipulate and enhance the  tripartite entanglement in a hybrid cavity-magnon optomechanical system. We first consider the single-parametric-driving case where either OPA or MPA is applied individually, for which we show how the parametric modulation can be utilized to achieve an enhanced tripartite entanglement among magnons, photons and phonons in the system. We then show that, when the OPA and MPA are simultaneously introduced, further enhancement of the  tripartite entanglement, as well as its robustness against thermal noise, can be achieved compared to the single-driving case. The underlying physics of such enhancement is verified and attributed to the constructive interference of the OPA and MPA when their phases are suitably matched. This also provides us a flexible way to manipulate the degree of entanglement by simply tuning the parameters of external parametric drivings. Our proposal thus establishes an efficient method to significantly enhance and flexibly regulate entanglement in hybrid quantum systems, thereby opening new possibilities for studying macroscopic
quantum phenomena and various entanglement-enabled quantum technologies, e.g., quantum networks \cite{wengerowsky2018entanglement,kimble2008quantum}.

\section{Theoretical model and quantum dynamics of the hybrid quantum system}\label{sec2}

\begin{figure*}
	\includegraphics[width=2\columnwidth]{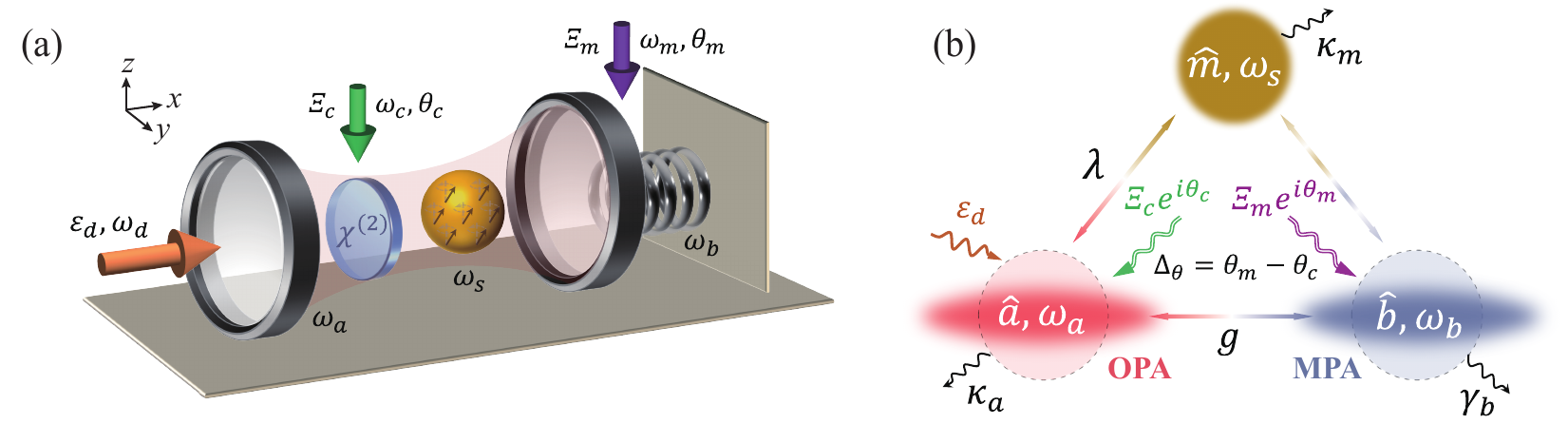}
	\caption{Schematic diagram (a) and physical model (b) of the parametrically-driven hybrid quantum system. The cavity supporting a photon mode $\hat{a}$ with frequency $\omega_a$ is driven by a coherent field with amplitude $\varepsilon_{d}$ and frequency $\omega_{d}$, inside which a nonlinear $\chi^{(2)}$ medium pumped by a nonlinear driving with amplitude $\it{\Xi}_c$, frequency $\omega_c$ and phase $\theta_c$ , and a magnetic sphere supporting a magnon mode $\hat{m}$ with frequency $\omega_s$ driven by a static magnetic ﬁeld (along the $z$ direction, not shown here) are placed. The mechanical oscillator supporting a mechanical mode $\hat{b}$ with frequency $\omega_b$ is pumped by a nonlinear driving of amplitude $\it{\Xi}_m$, frequency $\omega_m$ and phase $\theta_m$ that modulates the mechanical spring constant in time, resulting in the process of MPA. The cavity mode $\hat{a}$ is coupled to the mechanical mode $\hat{b}$ via radiation pressure with a rate of $g$, and is coupled to the magnon mode $\hat{m}$ via magnetic dipole interaction with a rate of $\lambda$. The phase difference between the OPA and MPA is $\Delta_{\theta}$.}
	\label{fig1}
\end{figure*}

We consider a hybrid cavity-magnon optomechanical system consisting of a cavity coupled to a mechanical oscillator and a spherical magnet, as schematically illustrated in Fig.~\ref{fig1}(a). We assume that the hybrid quantum system is driven by three external fields. Specifically, an external field of frequency $\omega_{c}$, amplitude $\it{\Xi}_{c}$ and phase $\theta_{c}$ is introduced to drive the nonlinear medium placed inside the cavity, which induces the process of OPA in the cavity \cite{PhysRevB.103.134409,PhysRevLett.95.140504,PhysRevA.84.053846}. The mechanical oscillator is electrically driven by a periodic pump of frequency $\omega_{m}$, amplitude $\it{\Xi}_{m}$ and phase $\theta_{m}$ that modulates the mechanical spring constant in time, generating the process of MPA \cite{PhysRevLett.67.699,PhysRevLett.107.213603,lemonde2016enhanced}. The cavity is simultaneously driven by a coherent field of frequency $\omega_{d}$ and amplitude $\varepsilon_{d}=\sqrt{2\kappa_{a}P/\hbar\omega_{d}}$, where $\kappa_{a}$ and $P$ denote the cavity decay rate and the input laser power, respectively. The magnet is assumed to be placed at the antinode of the magnetic field of the cavity and to interact with a static magnetic field along the $z$ direction that can align the magnetization of the magnet to $z$. In a frame rotating at the driving frequency $\omega_{d}$, the Hamiltonian of the hybrid system can be written as ($\hbar=1$)
\begin{equation}\label{eq1}
\begin{split}
\hat{H}=&\Delta_{a}\hat{a}^{\dagger}\hat{a}+\omega_{b}\hat{b}^{\dagger}\hat{b}+\Delta_{s}\hat{m}^{\dagger}\hat{m}-g\hat{a}^{\dagger}\hat{a}(\hat{b}^{\dagger}+\hat{b})\\
&\!+\!\lambda(\hat{a}^{\dagger}\hat{m}\!+\!\hat{m}^{\dagger}\hat{a})+i\it{\Xi}_{c}(\hat{a}^{\dagger\mathrm{2}}e^{-i\omega_c't+i\theta_c}\!-\!\hat{a}^{\mathrm{2}}e^{i\omega_c't-i\theta_c})\\
&\!+\!i\it{\Xi}_{m}(\hat{b}^{\dagger\mathrm{2}}e^{-i\omega_{m}t+i\theta_{m}}\!-\!\hat{b}^{\mathrm{2}}e^{i\omega_{m}t-i\theta_{m}})\!+\!i\varepsilon_{d}(\hat{a}^{\dagger}\!-\!\hat{a}),
\end{split}
\end{equation}
where $\hat{a}^{\dagger}$ ($\hat{a}$), $\hat{b}^{\dagger}$ ($\hat{b}$) and $\hat{m}^{\dagger}$ ($\hat{m}$) are the creation (annihilation) operators of the cavity mode with resonance frequency $\omega_{a}$, the mechanical mode with fundamental frequency $\omega_{b}$, and the Kittel magnon mode with eigenfrequency $\omega_{s}$, respectively. $\Delta_{a(s)}=\omega_{a(s)}-\omega_{d}$ is the frequency detuning between the cavity (magnon) and the coherent driving, and $\omega_{c}'=\omega_{c}-2\omega_{d}$ is the frequency detuning between the parametric and coherent drivings. In Eq.~(\ref{eq1}), the first three terms are free Hamiltonians for the cavity, mechanical and magnon modes. The $g$ term describes the optomechanical
interaction between the cavity and the mechanical oscillator, with $g$ being the single-photon optomechanical coupling strength. The $\lambda$ term denotes the magnetic dipole interaction between the cavity and the Kittel magnon with a tunable
coupling rate $\lambda$. The two quadratic terms with coefficients $\it{\Xi}_{c}$ and $\it{\Xi}_{m}$ describe the nonlinear (parametric) interactions induced by OPA and MPA, respectively. The last term denotes the interaction of the coherent driving field with the cavity.

Considering the dissipations of the cavity, mechanical and magnon modes, the dynamics of the hybrid system can be described by the following set of quantum Langevin equations (QLEs)
\begin{eqnarray}\label{QLEs}
\dot{\hat{a}}&=&-(\kappa_{a}+i\Delta_{a})\hat{a}+ig\hat{a}(\hat{b}+\hat{b}^{\dagger})-i\lambda\hat{m}+\varepsilon_{d}\nonumber\\&&+2\it{\Xi}_{c}\hat{a}^{\dagger}e^{-i(\omega_c't-i\theta_c)}+\sqrt{\mathrm{2}\kappa_{a}}\hat{a}^{in},\nonumber\\
\dot{\hat{b}}&=&-(\gamma_{b}+i\omega_{b})\hat{b}+ig\hat{a}^{\dagger}\hat{a}+2\it{\Xi}_{m}\hat{b}^{\dagger}e^{-i(\omega_mt-i\theta_m)}\\&&+\sqrt{\mathrm{2}\gamma_{b}}\hat{b}^{in},\nonumber\\
\dot{\hat{m}}&=&-(\kappa_{m}+i\Delta_{s})\hat{m}-i\lambda\hat{a}+\sqrt{\mathrm{2}\kappa_{m}}\hat{m}^{in}\nonumber,
\end{eqnarray}
where $\kappa_{a}$, $\gamma_{b}$ and $\kappa_{m}$ are the  intrinsic decay rates, and $\hat{a}^{in}$, $\hat{b}^{in}$ and $\hat{m}^{in}$ are the input noise operators, for the cavity, mechanical and magnon modes, respectively. These vacuum Gaussian noise operators have zero mean values and are characterized by the nonzero correlation functions \cite{PZoller_QN}
\begin{equation}
\begin{aligned}
	\left\langle\hat{a}^{i n}(t) \hat{a}^{i n, \dagger}\left(t^{\prime}\right)\right\rangle & =\left(\bar{n}_a+1\right)\delta\left(t-t^{\prime}\right), \\
	\left\langle\hat{a}^{in,\dagger}(t) \hat{a}^{in}\left(t^{\prime}\right)\right\rangle & =\bar{n}_a\delta\left(t-t^{\prime}\right), \\
	\left\langle\hat{b}^{i n}(t) \hat{b}^{i n, \dagger}\left(t^{\prime}\right)\right\rangle & =\left(\bar{n}_m+1\right) \delta\left(t-t^{\prime}\right), \\
	\left\langle\hat{b}^{\text {in, } \dagger}(t) \hat{b}^{\text {in }}\left(t^{\prime}\right)\right\rangle & =\bar{n}_m \delta\left(t-t^{\prime}\right), \\
	\left\langle\hat{m}^{i n}(t) \hat{m}^{i n, \dagger}\left(t^{\prime}\right)\right\rangle & =\left(\bar{n}_s+1\right)\delta\left(t-t^{\prime}\right),\\
	\left\langle\hat{m}^{in,\dagger}(t) \hat{m}^{in}\left(t^{\prime}\right)\right\rangle & =\bar{n}_s\delta\left(t-t^{\prime}\right),
\end{aligned}
\end{equation}
where $\bar{n}_{j}=\left[\exp(\hbar\omega_{j}/k_{B}T)-1\right] ^{-1}$ $(j=a,b,s)$ is the mean thermal occupation number for each mode, respectively, with $k_{B}$ being the Boltzmann constant and $T$ the bath temperature. 

QLEs (\ref{QLEs}) include the nonlinear optomechanical interaction between the cavity field and the mechanical oscillator, and thus are difficult to be directly solved. In order to solve these nonlinear dynamical equations, one can  linearize the system dynamics by expanding each operator as a sum of its steady-state mean value and a small quantum fluctuation around it under the condition of a strong cavity driving (giving a large cavity-field amplitude $\langle \hat{a}\rangle\gg1$), i.e., $\hat{O}(t)=\langle\hat{O}(t)\rangle+\delta \hat{O}(t)\ (\hat{O}=\hat{a}, \hat{b}, \hat{m})$. Substituting the operators into the QLEs (\ref{QLEs}), we then obtain the following time-evolutional dynamics of the steady-state mean values
\begin{eqnarray}\label{ssmve}
\langle\dot{\hat{a}}(t)\rangle&= & -(\kappa_{a}+i\Delta_{a})\langle\hat{a}(t)\rangle+i g\langle\hat{a}(t)\rangle(\langle\hat{b}(t)\rangle^*+\langle\hat{b}(t)\rangle)\nonumber\\
&&-i\lambda\langle\hat{m}(t)\rangle+2\it{\Xi}_c\langle\hat{a}(t)\rangle^* e^{-i(\omega_c' t-\theta_c)}+\varepsilon_{d},\nonumber\\
\langle\dot{\hat{b}}(t)\rangle&= & -(\gamma_b+i \omega_b)\langle\hat{b}(t)\rangle+2\it{\Xi}_m\langle\hat{b}(t)\rangle^* e^{-i\left(\omega_m t-\theta_m\right)}\\
&&+i g|\langle\hat{a}(t)\rangle|^2,\nonumber\\
\langle\dot{\hat{m}}(t)\rangle&= & -(\kappa_m+i \Delta_m)\langle\hat{m}(t)\rangle-i\lambda\langle\hat{a}(t)\rangle\nonumber,
\end{eqnarray}
and the linearized QLEs for the quantum fluctuations after neglecting the second-order fluctuation terms
\begin{equation}\label{lqleqs}
\begin{aligned}
\delta \dot{\hat{a}}=&-(\kappa_{a}+i\Delta) \delta \hat{a}+iG(\delta \hat{b}^{\dagger}+\delta \hat{b})-i\lambda\delta\hat{m}\\
&+2\it{\Xi}_c \delta \hat{a}^{\dagger} e^{-i(\omega_c't-\theta_c)}+\sqrt{\mathrm{2} \kappa_{a}}\hat{a}^{in}, \\
\delta \dot{\hat{b}}=&-(\gamma_{b}+i\omega_{b}) \delta \hat{b}+i(G \delta \hat{a}^{\dagger}+G^* \delta \hat{a})\\
&+2 \it{\Xi}_m \delta \hat{b}^{\dagger} e^{-i\left(\omega_m t-\theta_m\right)}+\sqrt{\mathrm{2} \gamma_b} \hat{b}^{i n},\\
\delta \dot{\hat{m}}=&-(\kappa_{m}+i\Delta_{s}) \delta \hat{m}-i\lambda\delta \hat{a}+\sqrt{\mathrm{2} \kappa_m} \hat{m}^{in},
\end{aligned}
\end{equation}
where $\Delta=\Delta_{a}-g(\langle\hat{b}(t)\rangle^*+\langle\hat{b}(t)\rangle)$ and $G=g\langle\hat{a}(t)\rangle$.

By defining the quadrature operators for the cavity, mechanical and magnon modes 
\begin{equation}
\begin{aligned}
\delta \hat{A}_{r} & =\frac{1}{\sqrt{2}}(\delta \hat{a}+\delta \hat{a}^{\dagger}), & \delta \hat{A}_{i} & =\frac{i}{\sqrt{2}}(\delta \hat{a}^{\dagger}-\delta \hat{a}), \\
\delta \hat{B}_{r} & =\frac{1}{\sqrt{2}}(\delta \hat{b}+\delta \hat{b}^{\dagger}), & \delta \hat{B}_{i}&=\frac{i}{\sqrt{2}}(\delta \hat{b}^{\dagger}-\delta \hat{b}),\\
\delta \hat{M}_{r} & =\frac{1}{\sqrt{2}}(\delta \hat{m}+\delta \hat{m}^{\dagger}), & \delta \hat{M}_{i}&=\frac{i}{\sqrt{2}}(\delta \hat{m}^{\dagger}-\delta \hat{m}),
\end{aligned}
\end{equation}
and the associated Hermitian input noise operators
\begin{equation}
\begin{aligned}
\hat{A}^{i n}_{r}&=\frac{1}{\sqrt{2}}(\hat{a}^{i n}+\hat{a}^{i n, \dagger}),  &\hat{A}^{i n}_{i}&=\frac{i}{\sqrt{2}}(\hat{a}^{i n, \dagger}-\hat{a}^{i n}), \\
\hat{B}^{i n}_{r}&=\frac{1}{\sqrt{2}}(\hat{b}^{i n}+\hat{b}^{i n, \dagger}),  &\hat{B}^{i n}_{i}&=\frac{i}{\sqrt{2}}(\hat{b}^{i n, \dagger}-\hat{b}^{i n}),\\
\hat{M}^{i n}_{r}&=\frac{1}{\sqrt{2}}(\hat{m}^{i n}+\hat{m}^{i n, \dagger}),  &\hat{M}^{i n}_{i}&=\frac{i}{\sqrt{2}}(\hat{m}^{i n, \dagger}-\hat{m}^{i n}),
\end{aligned}
\end{equation}
the linearized QLEs can be written as a compact form, i.e.,
\begin{equation}\label{cflQ}
\dot{\hat{u}}(t)=A(t) \hat{u}(t)+\hat{n}(t),
\end{equation}
where 
\begin{equation}
\begin{aligned}
\hat{u}(t)&=(\delta \hat{A}_{r}, \delta \hat{B}_{r}, \delta \hat{A}_{i}, \delta \hat{B}_{i}, \delta \hat{M}_{r}, \delta \delta \hat{M}_{i})^T, \\
\hat{n}(t)&=(\hat{A}^{i n}_{r}, \hat{A}^{i n}_{i},\hat{B}^{i n}_{r}, \hat{B}^{i n}_{i}, \hat{M}^{i n}_{r}, \hat{M}^{i n}_{i})^T,
\end{aligned}
\end{equation}
are the vectors of the quadrature fluctuation operators and the input noise operators, respectively, and the drift matrix $A(t)$ is given by	
\begin{equation}
\resizebox{1\hsize}{!}{$
A(t)=\left(\begin{array}{cccccc}
-\kappa_a+\zeta_{c}&\Delta-\Lambda_{c}&0 &\lambda&0&0\\
-\Delta-\Lambda_{c} &-\kappa_a-\zeta_{c}& -\lambda&0&2G&0\\
0&0&0&0&-\gamma_{b}+\zeta_{m} & \omega_{b}-\Lambda_{m}\\
2G&0&0&0&-\omega_{b}-\Lambda_{m}& -\gamma_{b}-\zeta_{m}\\
0 & \lambda & -\kappa_{m} & \Delta_{s} & 0 & 0 \\
-\lambda & 0 & -\Delta_s & -\kappa_{m} & 0 & 0
\end{array}\right),$}
\end{equation}
with $\zeta_{c}=2\it{\Xi}_{c}\cos(\omega_{c}'t-\theta_{c})$, $\zeta_{m}=2\it{\Xi}_{m}\cos(\omega_{m}t-\theta_{m})$, $\Lambda_{c}=2\it{\Xi}_{c}\sin(\omega_{c}'t-\theta_{c})$, and $\Lambda_{m}=2\it{\Xi}_{m}\sin(\omega_{m}t-\theta_{m})$.

One finds from the drift matrix $A(t)$ that the dynamics of the system could be affected by the parametric drivings ($\it{\Xi_{c}}$ and $\it{\Xi_{m}}$), which is relevant to the phase parameters ($\theta_{c}$ and $\theta_{m}$) of the drivings.  This can be interpreted more clearly by deriving the steady-state solutions $\langle \hat{O}(t)\rangle$ ($\hat{O}=\hat{a}, \hat{b}, \hat{m}$) via the Fourier series expansion \cite{PhysRevLett.103.213603,PhysRevA.100.043824,PhysRevA.81.033830,PhysRevA.86.053806,PhysRevA.98.023807}. For the case of applying a single optical parametric driving (assuming $\it{\Xi}_{m}=\mathrm{0}$), the driving strength $\it{\Xi}_{c}$ is small enough to avoid additional instabilities due to the parametric amplification and the system is far from optomechanical instabilities (true in the parameter regime we are interested in), one finds from Eq.~(\ref{ssmve}) that the mean values $\langle \hat{O}(t)\rangle$ may evolve toward an asymptotic periodic orbit with the same periodicity $2\pi/\omega_{c}'$ of the parametric driving. Also, as the effective driving amplitude $2\it{\Xi}_c\langle a(t)\rangle$ $\propto2\it{\Xi}_c\varepsilon_{d}$ (corresponding to the parametric driving) is much weaker than the cavity driving field $\varepsilon_{d}$, we can derive the steady-state solutions of Eq.~(\ref{ssmve}) to first order in $\varepsilon_{s}=2\it{\Xi}_c\varepsilon_{d}$ \cite{PhysRevLett.103.213603,PhysRevA.100.043824,PhysRevA.81.033830,PhysRevA.86.053806,PhysRevA.98.023807}, that is, we find $t\rightarrow\infty$ limit of the solutions:
\begin{equation}\label{eqA1}
	\langle \hat{O}(t)\rangle= O_{0}+ \varepsilon_s e^{-i\omega_{c}' t}O_{+}+\varepsilon_s^{*} e^{i\omega_{c}' t}O_{-},
\end{equation}
with $O_{0,\pm}$ being time-independent coefficients. Substituting Eq.~(\ref{eqA1}) into Eq.~(\ref{ssmve}) and equating coefficients of
terms proportional to $e^{i0\omega_{c}' t}$, $e^{-i\omega_{c}' t}$ and $e^{i\omega_{c}' t}$, respectively, a set of recursive formulas for the coefficients $O_{0}$, $O_{+}$ and $O_{-}$ can then be obtained. It is readily to get the coefficients
\begin{eqnarray}\label{eqA2}
a_{0}&=&\frac{i\lambda m_{0}-\varepsilon_{d}}{i g (b_{0}+b_{0}^{*})-\kappa_{a}-i\Delta_{a}},\nonumber \\
b_{0}&=&\frac{ig|a_{0}|^2}{\gamma_{b}+i\omega_{b}},\\
m_{0}&=&\frac{-i\lambda a_{0}}{\kappa_{m}+i\Delta_{m}}\nonumber.
\end{eqnarray}
The expressions for $a_{\pm}$ can be given by combining the following two formulas
\begin{equation}
\begin{split}	a_{+}=&\left[\mathcal{A}-ig(b_{0}+b_{0}^{*})+\frac{g^2|a_{0}|^2}{\mathcal{B}}+\frac{\lambda^{2}}{\mathcal{C}}-\frac{g^{2}a_{0}^{2}}{\mathcal{E}^{*}}\right]^{-1}\\&\times g^{2}a_{0}^2\left(\frac{1}{\mathcal{E}^{*}}-\frac{1}{\mathcal{B}}\right)a_{-}^{*}+\frac{2\it{\Xi}_{c}a_{\mathrm{0}}^{*}}{\varepsilon_{s}}e^{i\theta_{c}},
\end{split}
\end{equation}
and
\begin{equation}\label{eqA4}
\begin{split}
a_{-}=&\left[\mathcal{D}-ig(b_{0}+b_{0}^{*})+\frac{g^2|a_{0}|^2}{\mathcal{E}}-\frac{g^2a_{0}^2}{\mathcal{B}^*}+\frac{\lambda^{2}}{\mathcal{F}}\right]^{-1}\\&\times g^2a_{0}^2(\frac{1}{\mathcal{E}}-\frac{1}{\mathcal{B}^*})a_{+}^{*},
\end{split}
\end{equation}
where $\mathcal{A}=\kappa_{a}+i\Delta_{a}-i\omega_{c}'$, $\mathcal{B}=\gamma_{b}+i\omega_{b}-i\omega_{c}'$, $\mathcal{C}=\kappa_{m}+i\Delta_{m}-i\omega_{c}'$, $\mathcal{D}=\kappa_{a}+i\Delta_{a}+i\omega_{c}'$, $\mathcal{E}=\gamma_{b}+i\omega_{b}+i\omega_{c}'$, and $\mathcal{F}=\kappa_{m}+i\Delta_{m}+i\omega_{c}'$.
After simple algebra, we have
\begin{equation}\label{eqA5}
\begin{split}
a_{+}=&\frac{2\it{\Xi}_{c}a_{\mathrm{0}}^{*}e^{i\theta_{c}}}{\varepsilon_{s}} \left[\mathcal{A}-ig(b_{0}+b_{0}^{*})+\frac{g^2|a_{0}|^2}{\mathcal{B}}+\frac{\lambda^{2}}{\mathcal{C}}-\frac{g^{2}a_{0}^{2}}{\mathcal{E}^{*}}\right.\\
&\left.-\frac{g^4|a_{0}|^4(\frac{1}{\mathcal{E}}-\frac{1}{\mathcal{B}^*})}{\mathcal{D}^{*}+ig(b_{0}+b_{0}^{*})+\frac{g^2|a_{0}|^2}{\mathcal{E}^{*}}-\frac{g^2a_{0}^2}{\mathcal{B}}+\frac{\lambda^{2}}{\mathcal{F}^{*}}}\right]^{-1}.
\end{split}
\end{equation}
The explicit expression for $a_{-}$ can then be written out using Eq.~(\ref{eqA4}), which is not given here for brevity. Also, we note that the expressions for $b_{\pm}$ and $m_{\pm}$ are not explicitly given since they are unrelated to our discussion. By substituting $a_{0}$, $a_{+}$ and $a_{-}$ into Eq.~(\ref{eqA1}), the analytical approximation for the steady-state cavity-field mean value $\langle\hat{a}(t)\rangle$ can be obtained. With this analytical approximation, it is found that the cavity-field amplitude $|\langle a(t)\rangle|$ can be periodically modulated by the OPA (see Appendix \ref{appenA} for details). It is also found, both analytically and numerically, that $|\langle a(t)\rangle|$ is insensitive to the parametric phase $\theta_{c}$, featuring an unchanged maximal value of $|\langle a(t)\rangle|$ with varying $\theta_{c}$. This can be interpreted by working in a rotated frame with respect to half $\theta_{c}$, where the phase $\theta_{c}$ is shifted completely to the coherent driving and thus produce only an overall phase shift. In contrast, in the simultaneous presence of both OPA and MPA, $|\langle a(t)\rangle|$ can be modulated by the phase difference $\Delta_{\theta}$ between the two parametric drivings (see Appendix \ref{appenA}), exhibiting a behavior akin to a typical interference effect. As we will demonstrate below, such an interference enables the modulation and enhancement (under proper phase matching) of the entanglement.

Because of the zero-mean Gaussian nature of the quantum noises and the linearized dynamics of the quantum fluctuations that we are dealing with, the steady state of the quantum fluctuations is a continuous variable tripartite Gaussian state that can be fully characterized by a
$6 \times 6$ covariance matrix (CM) $\mathcal{V}(t)$, whose matrix elements are defined via
\begin{equation}\label{CMEq}
\mathcal{V}_{k, l}(t)\equiv\left\langle\hat{u}_k(t) \hat{u}_l(t)+\hat{u}_l(t) \hat{u}_k(t)\right\rangle / 2.
\end{equation}
Then, combining Eqs.~(\ref{cflQ}) and (\ref{CMEq}), we have an equation of motion of the CM  $\mathcal{V}(t)$ \cite{PhysRevLett.103.213603}
\begin{equation}\label{CMeq}
\dot{\mathcal{V}}(t)=A(t) \mathcal{V}(t)+\mathcal{V}(t) A^T(t)+D,
\end{equation}
where $D=\mathrm{diag}(\kappa_{a}(2\bar{n}_{a}+1),\kappa_{a}(2\bar{n}_{a}+1),\gamma_{b}(2\bar{n}_{b}+1),\gamma_{b}(2\bar{n}_{b}+1),\kappa_{m}(2\bar{n}_{m}+1),\kappa_{m}(2\bar{n}_{a}+1)) $ is the diffusion matrix and is defined via $D_{k, l} \delta\left(s-s^{\prime}\right)\equiv\left\langle\hat{n}_k(s) \hat{n}_l\left(s^{\prime}\right)+\hat{n}_l\left(s^{\prime}\right) \hat{n}_k(s)\right\rangle / 2$. For the proposed system, it should be stable in the long-time limit such that a unique steady-state solution of the system can be obtained. We examine (in the numerical simulations below) the stability of system by using the Routh-Hurwitz criterion which states that the system is stable if the real parts of all the eigenvalues of the drift matrix $A$ are negative.

\section{Enhancing the tripartite entanglement via phase-matched parametric amplification}\label{sec3}
In this section, we will show how to manipulate and enhance the tripartite entanglement using phase-matched parametric amplification. To this end, we first introduce quantitative measures for tripartite entanglement, i.e., the \textit{minimum} residual contangle, based on the CM derived above. Then, we verify, with detailed numerical calculations, the feasibility of using a single parametric driving (i.e., OPA or MPA modulation) to enhance the entanglement, and more importantly, we finally show the synergistic enhancement and flexible modulation of the entanglement with phase-matched OPA and MPA. 

\subsection{Quantitative measure of the tripartite entanglement}
To investigate the tripartite entanglement of the hybrid system, we adopt the quantitative measure of the \textit{minimum} residual contangle $R_{\tau}^{\mathrm{min}}$. To this end, we first introduce the logarithmic negativity $E_{N}$ that is widely adopted for quantitatively measuring the bipartite entanglement \cite{PhysRevA.65.032314,PhysRevLett.95.090503,PhysRevA.70.022318}, which is defined as 
\begin{equation}\label{EN}
E_N \equiv \max \left[0,-\ln 2 \tilde{v}_{-}\right],
\end{equation}
where $\tilde{v}_{-}=\min \operatorname{eig}|i \Omega_2 \tilde{\mathcal{V}}_4|$ (with the symplectic matrix $\Omega_2=\oplus_{j=1}^2 i \sigma_y$ and $\sigma_y$ the $y$-Pauli matrix) is the minimum symplectic eigenvalue of the CM $\tilde{\mathcal{V}}_4=\mathcal{P}_{1 \mid 2} \mathcal{V}_4 \mathcal{P}_{1 \mid 2}$, where $\mathcal{V}_4$ is the $4\times4$ CM of two subsystems, obtained by removing in $\mathcal{V}$ the rows and columns of the uninteresting mode, and $\mathcal{P}_{1 \mid 2}\equiv\operatorname{diag}\left[1,-1,1,1\right]$. The \textit{minimum} residual contangle $R_{\tau}^{\mathrm{min}}$, which provides a \textit{bona fide} quantification of continuous-variable tripartite entanglement, is defined as \cite{Adesso_2007,Adesso_2006}
\begin{equation}
R_{\tau}^{\mathrm{min}}=\mathop {\min}\limits_{(r,s,t)}[E_{\tau}^{r\mid st}-E_{\tau}^{r\mid s}-E_{\tau}^{r\mid t}],
\end{equation}
where $(r,s,t)\in\left\lbrace a,b,m\right\rbrace$ denotes all possible permutations of the three-mode indexes. $E_{\tau}^{u\mid v}$ is the contangle of subsystems of $u$ ($u$ contains one mode) and $v$ ($v$ contains one or
two modes), which can be defined by a proper entanglement
monotone, e.g., the squared logarithmic negativity. When calculating the one-mode-vs-one-mode contangle $E_{\tau}^{r\mid s}$ or $E_{\tau}^{r\mid t}$, its definition can be directly used based on Eq.~(\ref{EN}), i.e., $E_{\tau}\equiv[E_{N}]^{2}$. As for the case of one-mode-vs-two-modes, one can obtain $E_{r \mid s t}$ simply by replacing $\Omega_2$ in Eq.~(\ref{EN}) with $\Omega_3=\oplus_{j=1}^3$ $i \sigma_y$, and $\tilde{\mathcal{V}}_4$ with $\tilde{\mathcal{V}}=\mathcal{P}_{r \mid s t} \mathcal{V} \mathcal{P}_{r \mid s t}$, where $\mathcal{P}_{r \mid st}\equiv\operatorname{diag}(1,-1,1,1,1,1)$. In terms of $E_{s \mid r t}$ and $E_{t \mid s r}$, the corresponding matrices are $\mathcal{P}_{s \mid rt}\equiv\operatorname{diag}(1,1,1,-1,1,1)$ and $\mathcal{P}_{t \mid sr}\equiv\operatorname{diag}(1,1,1,1,1,-1)$, respectively. The residual contangle is required to satisfy the monogamy condition according to the Coffman-Kundu-Wootters monogamy inequality \cite{PhysRevA.61.052306}, i.e., $E_{\tau}^{r\mid st}-E_{\tau}^{r\mid s}-E_{\tau}^{r\mid t}\ge0$, which implies that the \textit{genuine} tripartite entanglement is present in the system provided $R_{\tau}^{\mathrm{min}}>0$.

The differential equations (\ref{ssmve}) for the classical mean values indicate that the system may evolve periodically, which has been verified by the results in, e.g., Fig.~\ref{fig4}(b). This suggests that the long-time dynamics of the system follows the initial one, and thus the drift matrix $A(t)$ is $\tau$ periodic, satisfying $A(t+\tau)=A(t)$.
Meanwhile, since Eq.~(\ref{CMeq}) is a linear differential equation with periodic
coefficients, $\mathcal{V}(t)$ can persevere the same periodicity in the long time limit, i.e., $\mathcal{V}(t+\tau)=\mathcal{V}(t)$. The \textit{minimum} residual contangle $R_{\tau}^{\mathrm{min}}$ is therefore  $\tau$ periodic. In this case, we hereafter quantify the tripartite entanglement by its maximum in one period, namely,
\begin{equation}
\begin{aligned}
R_{\tau,\mathrm{max}}^{\mathrm{min}}&=\mathop {\max}\limits_{\tau}\left[R_{\tau}^{\mathrm{min}}(t)\right].
\end{aligned}
\end{equation}

\begin{figure*}
	\includegraphics[width=1.6\columnwidth]{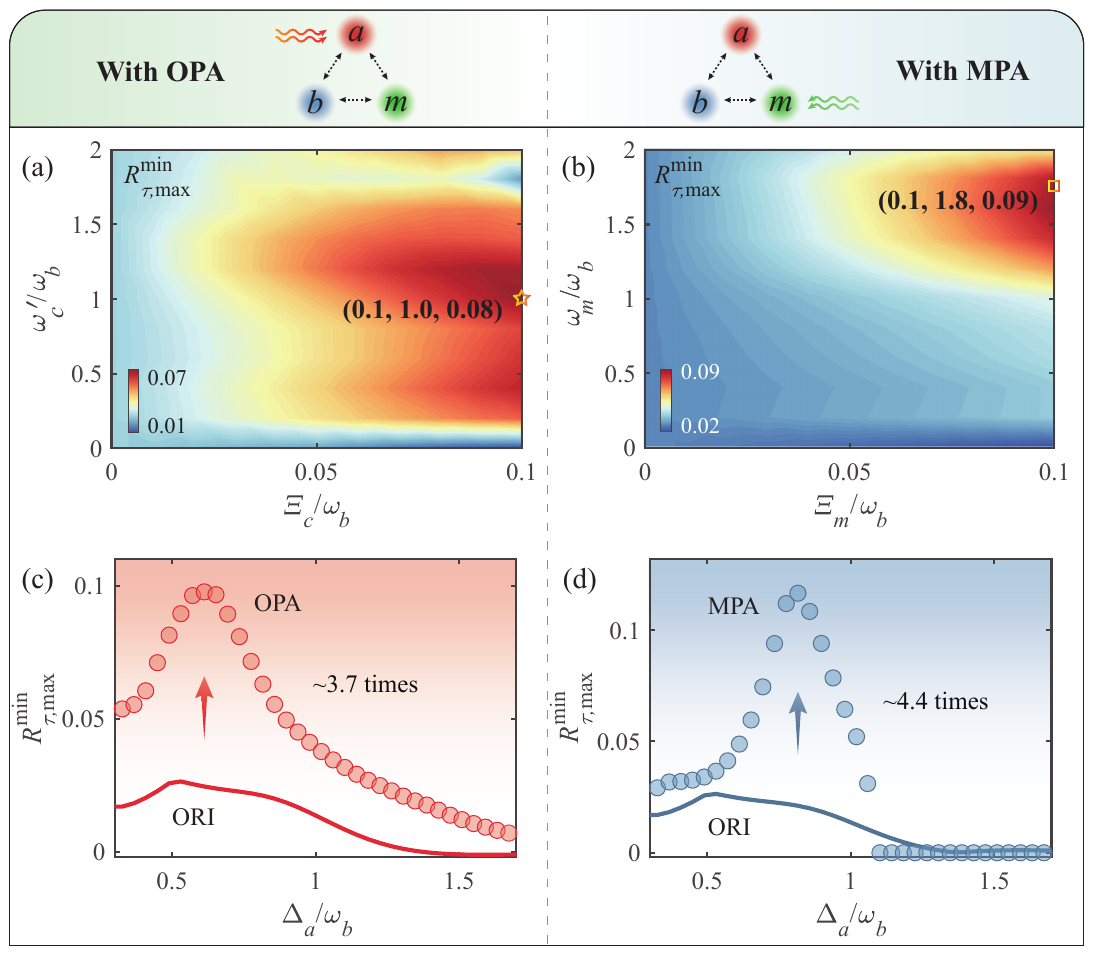}
	\caption{Tripartite entanglement under single parametric amplification. (a) The photon-phonon-magnon tripartite entanglement characterized by the minimum residual contangle $R_{\tau,\mathrm{max}}^{\mathrm{min}}$ versus the scaled optical parametric frequency $\omega_c'/\omega_b$ and amplitude $\it{\Xi}_c/\omega_b$. (b) $R_{\tau,\mathrm{max}}^{\mathrm{min}}$ versus the mechanical parametric frequency $\omega_m/\omega_b$ and amplitude $\it{\Xi}_m/\omega_b$. (c),(d) $R_{\tau,\mathrm{max}}^{\mathrm{min}}$ versus the scaled cavity detuning $\Delta_{a}/\omega_{b}$ without (solid curves) and with OPA or MPA (circles), showing the enhancement of tripartite entanglement by OPA or MPA. Here, ``ORI'' denotes the case with neither OPA nor MPA. The parameters used here are $\Delta_s/\omega_b=-1$, $g/\omega_b=5\times10^{-6}$, $\lambda/\omega_b=0.5$,  $\kappa_a/\omega_b=\kappa_m/\omega_b=0.2$, $\gamma_b/\omega_b=10^{-5}$, $\theta_{c}=\theta_{m}=0$, $T=0.01$ K, in (a) and (b), $\Delta_a/\omega_b=0.73$, in (c), $\it{\Xi}_c/\omega_b=\mathrm{0.1}$, $\omega_c'/\omega_b=1.0$, and in (d), $\it{\Xi}_m/\omega_b=\mathrm{0.1}$, $\omega_m/\omega_b=1.8$.}
	\label{fig3}
\end{figure*}

\subsection{Entanglement enhancement with single parametric amplification}\label{sec3A}
From the linearized Langevin equations (\ref{lqleqs}), a linearized Hamiltonian for the hybrid system described originally by Eq.~(\ref{eq1}) can be readily obtained, which is given by
\begin{equation}\label{eq16}
\begin{split}
\hat{H}'=&\Delta\hat{a}^{\dagger}\hat{a}+\omega_{b}\hat{b}^{\dagger}\hat{b}+\Delta_{s}\hat{m}^{\dagger}\hat{m}-G(\hat{a}^{\dagger}+\hat{a})(\hat{b}^{\dagger}+\hat{b})\\
&\!+\!\lambda(\hat{a}^{\dagger}\hat{m}\!+\!\hat{m}^{\dagger}\hat{a})+i\it{\Xi}_{c}(\hat{a}^{\dagger\mathrm{2}}e^{-i\omega_c't+i\theta_c}\!-\!\hat{a}^{\mathrm{2}}e^{i\omega_c't-i\theta_c})\\
&\!+\!i\it{\Xi}_{m}(\hat{b}^{\dagger\mathrm{2}}e^{-i\omega_{m}t+i\theta_{m}}\!-\!\hat{b}^{\mathrm{2}}e^{i\omega_{m}t-i\theta_{m}}).
\end{split}
\end{equation}
In the parameter regime $\Delta\approx\omega_{b}$, the optomechanical beam-splitter interactions characterized by $\hat{a}^\dagger\hat{b}+\hat{b}^\dagger\hat{a}$ in Eq.~(\ref{eq16}) are nearly resonant, which can significantly cool the mechanical mode to the ground state (which is prerequisite for preparing any quantum states). In contrast, for $\Delta\approx-\omega_{b}$, the parametric down-conversion processes characterized by $\hat{a}^\dagger\hat{b}^\dagger+\hat{b}\hat{a}$ are dominant, which are known to generate optomechanical entanglement \cite{PhysRevLett.98.030405,PhysRevA.78.032316}. It is intuitive that the best regime for achieving a strong optomechanical entanglement is $\Delta\approx-\omega_{b}$, where the parametric-down conversion process is resonant. However, achieving sufficient amount of entanglement at the steady state 
is seriously hindered by instability in the parametric-down-conversion regime \cite{PhysRevA.78.032316}, and one has to work in the opposite beam-splitter regime of $\Delta\approx\omega_{b}$. As we will show below, the parameter regime for finding significant tripartite entanglement in our system is shifted away from the beam-splitter regime $\Delta\approx\omega_{b}$, which is due to the fact that the magnon-photon interaction effectively shifts the cavity detuning $\Delta_{a}$. Once the optomechanical entanglement is generated, it can then be distributed to the magnon provided that the effective magnon frequency matches the Stokes sideband, i.e., $\Delta_{s}=-\omega_{b}$, thereby resulting in the magnon-phonon and photon-magnon entanglement. Having obtained all bipartite entanglement of the system, a \textit{genuine} tripartite entanglement can be achieved. In the following, we proceed to examine numerically the effect of introducing OPA and MPA on the tripartite entanglement.

\begin{figure*}
	\includegraphics[width=2\columnwidth]{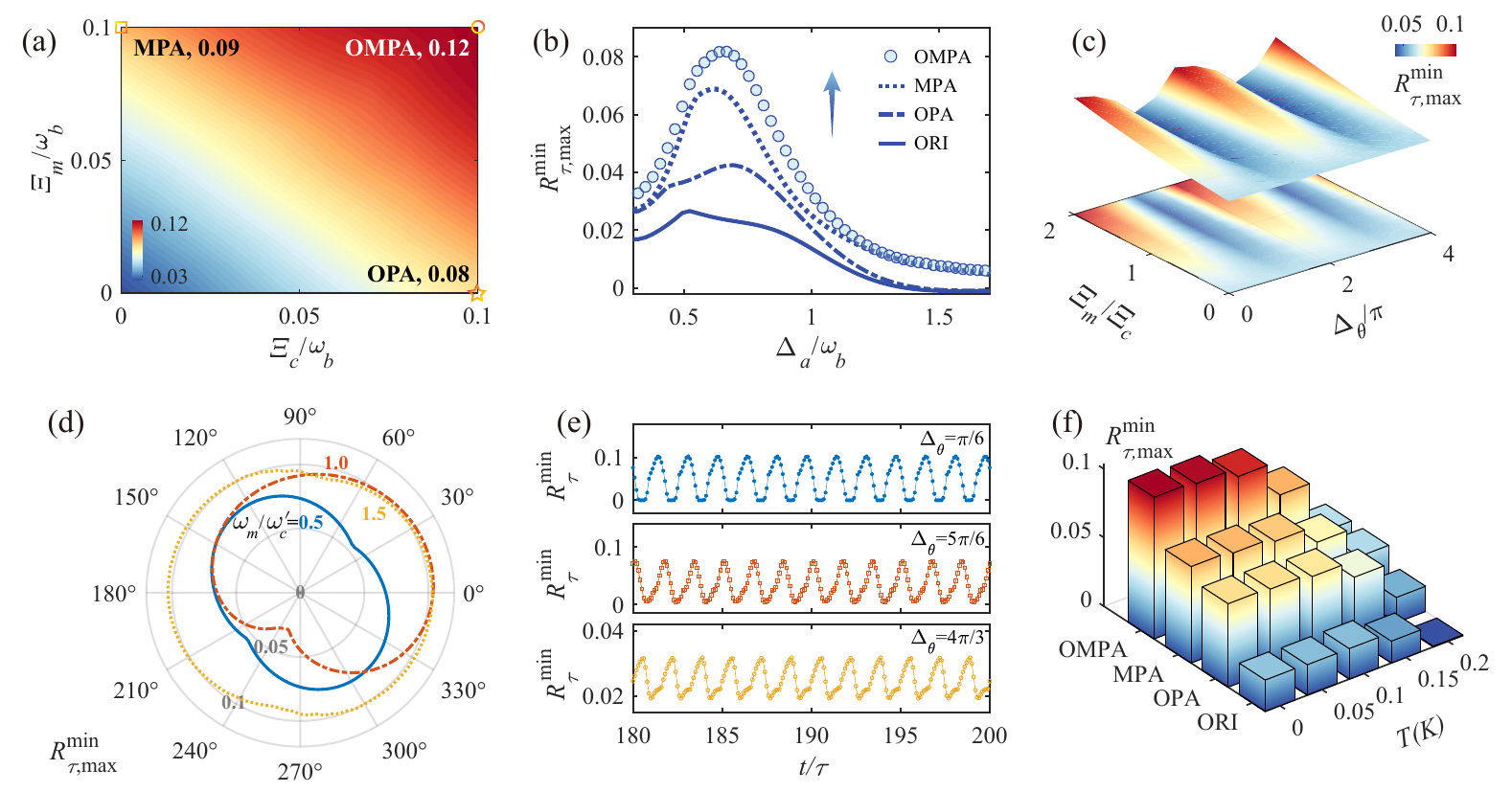}
	\caption{Enhancement of the tripartite entanglement under phase-matched parametric amplification. (a) The minimum residual contangle $R_{\tau,\mathrm{max}}^{\mathrm{min}}$ versus the scaled parametric amplitudes $\it{\Xi}_m/\omega_b$ and $\it{\Xi}_c/\omega_b$, for a phase difference of $\Delta_{\theta}=0$. The hollow square in the top left and the hollow pentagram in the bottom right correspond, respectively, to the markers in Figs.~\ref{fig3}(a) and (b). (b) $R_{\tau,\mathrm{max}}^{\mathrm{min}}$ versus the scaled cavity detuning $\Delta_{a}/\omega_{b}$ without (solid curves), with OPA (dash-dotted curves), MPA (dotted curves), and OMPA (circles) modulations, for $\Delta_{\theta}=0$. (c) $R_{\tau,\mathrm{max}}^{\mathrm{min}}$ versus the amplitude ratio  $\it{\Xi}_m/\it{\Xi}_c$ and the phase difference $\Delta_{\theta}$, showing a periodic modulation of the entanglement by varying $\Delta_{\theta}$. (d) $R_{\tau,\mathrm{max}}^{\mathrm{min}}$ versus $\Delta_{\theta}$ in a polar coordinate for different values of frequency ratio $\omega_{m}/\omega_{c}'$ and $\it{\Xi}_m/\it{\Xi}_c=\mathrm{2}$. (e) Time evolution of $R_{\tau}^{\mathrm{min}}$ for different values of $\Delta_{\theta}$, $\it{\Xi}_m/\it{\Xi}_c=\mathrm{2}$ and $\omega_{m}/\omega_{c}'=1$. (f) Comparison of $R_{\tau,\mathrm{max}}^{\mathrm{min}}$ under different temperatures $T$ without and with three different parametric modulations, showing the strongest robustness against thermal noise of the OMPA modulation. The parameters are $\Delta_a/\omega_b=0.73$, in (a), $\omega_c'/\omega_b=1.0$, $\omega_m/\omega_b=1.8$, in (b-f), $\it{\Xi}_c/\omega_b=\it{\Xi}_m/\omega_b=\mathrm{0.05}$, $\omega_c'/\omega_b=\omega_m/\omega_b=1.2$. Other common parameters are the same as in Fig.~\ref{fig3}.} 
	\label{fig5}
\end{figure*}

In our numerical calculations, we have selected the experimentally feasible parameters: $\omega_{b}/2\pi=10$ MHz, $\omega_{a}/2\pi=10$ GHz, $g/2\pi=50$ Hz, $\gamma_{b}/2\pi=100$ Hz, $\kappa_{a}/2\pi=2$ MHz and $T=0.01$ K \cite{toth2017dissipative,PhysRevApplied.20.024039}; $\lambda/2\pi=5$ MHz, $\kappa_{m}/2\pi=2$ MHz \cite{zhang2016cavity,PhysRevLett.129.243601,PhysRevApplied.20.024039}. The maxima of $\left\lbrace \it{\Xi}_c,\it{\Xi}_m\right\rbrace $ and $\left\lbrace\omega_{c}',\omega_{m}\right\rbrace $ are set, respectively, as $2\pi\times1$ MHz and $2\pi\times20$ MHz in the numerical simulations, which could be experimentally achieved \cite{PRXQuantum.5.020306,PhysRevLett.111.207203}. In addition, $\left\lbrace \omega_{d},\omega_{s}\right\rbrace/2\pi \approx10$ GHz and $\omega_{c}/2\pi\approx20$ GHz are chosen according to the specific parameter conditions with respect to $\Delta_{a}$ and $\Delta_{s}$ considered in the simulations, which are also accessible in current experiments \cite{PRXQuantum.5.020306,PhysRevLett.123.107701}. In Fig.~\ref{fig3}(a) we plot the \textit{minimum} residual contangle $R_{\tau,\mathrm{max}}^{\mathrm{min}}$ versus the scaled parametric-driving detuning $\omega_{c}'/\omega_{b}$ and amplitude $\it{\Xi}_c/\omega_b$ at a fixed value of the scaled cavity detuning $\Delta_{a}/\omega_{b}=0.73$, for the system under the OPA only. A prominent increase in $R_{\tau,\mathrm{max}}^{\mathrm{min}}$ with increasing $\it{\Xi}_c/\omega_b$ for proper $\omega_c'/\omega_b$ reveals that the tripartite entanglement of the system can be greatly enhanced by the OPA. For $\it{\Xi}_c/\omega_b=\mathrm{0.1}$ and $\omega_{c}'/\omega_{b}=1.0$ (marked by a hollow pentagram), $R_{\tau,\mathrm{max}}^{\mathrm{min}}$ is increased to a maximum of about 0.08. Moreover, by slightly tuning the scaled cavity detuning $\Delta_{a}/\omega_{b}$ to 0.61, as depicted in Fig.~\ref{fig3}(c), $R_{\tau,\mathrm{max}}^{\mathrm{min}}$ can be further enhanced to near 0.1, which is about 3.7 times higher than its maximum for the case without OPA. This enhancement is physically a consequence of the OPA increasing the photon populations in the cavity and changing the photon statistics of the cavity mode, thereby directly improving the photon-phonon and photon-magnon interactions. Such enhancement in the coupling strengths boosts the cooling of the mechanical mode and the generation as well as transfer of the bipartite entanglement, and consequently, produces a considerable gain for the tripartite entanglement when the system working with appropriate parameters. In a similar manner, we predict that applying the MPA could be also effective for improving the photon-phonon interaction and therefore the tripartite entanglement of the system. We verify this in Fig.~\ref{fig3}(d), where more than 4 times enhancement in $R_{\tau,\mathrm{max}}^{\mathrm{min}}$ is within reach for $\it{\Xi}_m/\omega_b=\mathrm{0.1}$, $\omega_m/\omega_b=1.8$ and $\Delta_{a}/\omega_{b}=0.81$.

\subsection{Improvement of the enhanced entanglement by synergizing parametric amplifications}

\begin{figure}
	\includegraphics[width=0.97\columnwidth]{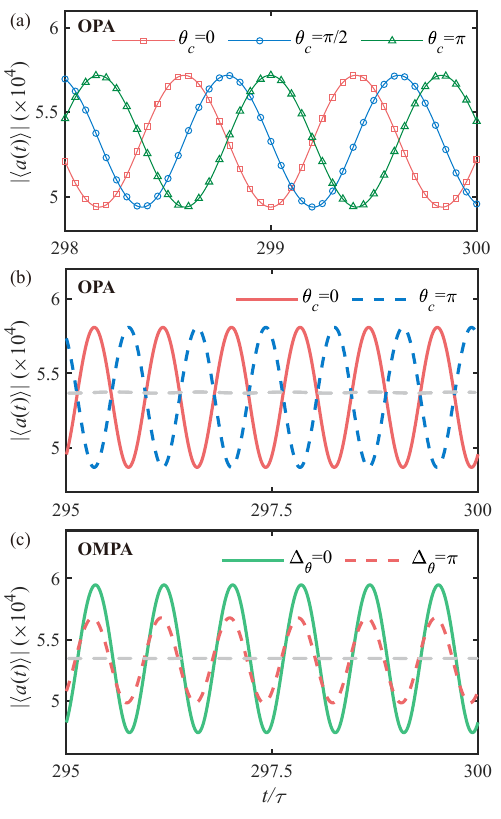}
	\caption{(a) Analytical approximation of the long-time dynamics of the cavity-field amplitude $|\langle a(t)\rangle|$ under OPA, with $\tau\equiv2\pi/\omega_b$. (b),(c) Numerical result of the long-time dynamics of $|\langle a(t)\rangle|$ under (b) OPA and (c) both OPA and MPA, revealing periodical evolution of the system under the parametric modulation. For the case of individually applying OPA, the maximal value of $|\langle a(t)\rangle|$ remains unchanged with varying the parametric phase $\theta_{c}$; whereas for the simultaneous application of both OPA and MPA, the maximal $|\langle a(t)\rangle|$ can be modulated by the phase difference $\Delta_\theta$. The gray dashed curves correspond to the cases without any parametric drivings. ``OMPA'' in (d) means the simultaneous existence of OPA and MPA. The parameters used here are $\Delta_a/\omega_b=1$, $\Delta_s/\omega_b=-1$, $g/\omega_b=5\times10^{-6}$, $\lambda/\omega_b=0.5$,  $\varepsilon_d/\omega_b=6\times10^4$, $\kappa_a/\omega_b=\kappa_m/\omega_b=0.2$, $\gamma_b/\omega_b=10^{-5}$, $T=0$, in (a)-(c),  $\it{\Xi}_c/\omega_b=\mathrm{0.01}$, $\omega_c'/\omega_b=1.2$, and in (c), $\it{\Xi}_m/\omega_b=\mathrm{0.01}$, $\omega_m/\omega_b=1.2$.}
	\label{fig4}
\end{figure}

As discussed above, introducing OPA or MPA into the system provides a feasible way to enhance and manipulate the tripartite entanglement. Perhaps more interesting is the ability of our scheme to further improve the entanglement by exploiting the synergistic effect of OPA and MPA. This relies on a suitable phase-matching between the two parametric drivings that periodically modulates the strength of entanglement, resembling an interference effect.  Figure \ref{fig5} summarizes the numerical results of the behavior of the tripartite entanglement under OPA and MPA. Panel (a) shows the dependence of the \textit{minimum} residual contangle $R_{\tau,\mathrm{max}}^{\mathrm{min}}$ on the scaled parametric amplitudes $\it{\Xi}_c/\omega_b\in\left[\mathrm{0,0.1}\right]$,  $\it{\Xi}_m/\omega_b\in\left[\mathrm{0,0.1}\right]$ and $\Delta_{\theta}=0$, with the other parameters: $\omega_{c}'/\omega_{b}=1$, $\omega_{m}/\omega_{b}=1.8$ and $\Delta_{a}/\omega_{b}=0.73$. Here, the parameters are specifically selected to accord with Fig.~\ref{fig3}, so that the maximal values of $R_{\tau,\mathrm{max}}^{\mathrm{min}}$ (within the same parameter regime as Fig.~\ref{fig3}) under both and individual parametric drivings can be directly compared. 
The result shows that $R_{\tau,\mathrm{max}}^{\mathrm{min}}$ is raised to its maximum of about 0.12 for $\it{\Xi}_c/\omega_b=\it{\Xi}_m/\omega_b=\mathrm{0.1}$ where both parametric drivings are present simultaneously, which is higher than the cases of individually applying OPA ($R_{\tau,\mathrm{max}}^{\mathrm{min}}=0.08$) or MPA ($R_{\tau,\mathrm{max}}^{\mathrm{min}}=0.09$), revealing an enhanced tripartite entanglement by the interplay of OPA and MPA.
In addition, the enhancement of entanglement can be regulated flexibly by various parameters of the system such as the cavity detuning (see panel (b)), the phase difference and amplitudes (see panel (c)), as well as the frequencies (see panel (d)) of the parametric drivings, allowing us to find a regime of parameters where the entanglement is maximized. As depicted in Fig.~\ref{fig5}(b), $R_{\tau,\mathrm{max}}^{\mathrm{min}}$, with and without parametric amplifications, are compared by varying $\Delta_{a}/\omega_{b}$. Hereafter, the parametric amplitudes are set as $\it{\Xi}_c/\omega_b=\it{\Xi}_m/\omega_b=\mathrm{0.05}$ to maintain the stability of system. The combination of OPA and MPA exhibits the strongest entanglement compared to applying just one of the parametric drivings, for most values of $\Delta_{a}/\omega_{b}$. The underlying physics of such an enhancement is as follows. When the OPA and MPA with suitably matched phases (e.g., $\Delta_{\theta}=0$) are both present, the optomechanical coupling can be enhanced simultaneously by both optical and mechanical parametric drivings, resulting in a further increase in the maximal $|\langle a(t) \rangle|$ compared with the individual cases (as demonstrated in Fig. \ref{fig1}(e)). As the strength of the photon-phonon coupling (which is crucial for generating entanglement) depends directly on the value of $|\langle a(t) \rangle|$, the tripartite entanglement gets further improved.
Notice in Fig.~\ref{fig5}(c) that the strength of entanglement is regulated periodically, as expected, by the phase difference between the OPA and MPA, and a further enhancement of entanglement can be achieved by increasing the ratio of amplitudes of the parametric drivings $\it{\Xi}_m/\it{\Xi}_c$.  Figure \ref{fig5}(e) plots the time evolution of $R_{\tau}^{\mathrm{min}}$ for $\omega_{m}/\omega_{c}'=1$ and different values of $\Delta_{\theta}$. The periodic evolution and amplitude modulation (by $\Delta_{\theta}$) of $R_{\tau}^{\mathrm{min}}$ is consistent with the discussion on the cavity-field amplitude $|\langle a(t)\rangle|$ shown in Fig.~\ref{fig4}(c), and such a comparison in the entanglement dynamics demonstrates more clearly the nature and property of the phase-difference-modulated entanglement in our scheme. In view of the effect of the thermal noise on the entanglement, we summarize $R_{\tau,\mathrm{max}}^{\mathrm{min}}$ under various bath temperature $T$, considering four cases without and with (different) parametric drivings, as shown in Fig.~\ref{fig5}(f). For a same temperature, $R_{\tau,\mathrm{max}}^{\mathrm{min}}$ for the case of OMPA exhibits a much higher value than that of other three cases, implying an enhanced robustness against thermal noise under the interplay of OPA and MPA.

\section{Discussion and Conclusion}\label{sec4}
A range of different approaches can be envisaged for experimental realization of our proposed setup. As an example of solid-state system, we suggest to couple a low damping magnetic insulator such as yttrium iron garnet (YIG) to a microwave cavity with a nonlinear medium that supports microwave parametric down-conversion \cite{PhysRevLett.95.140504,PhysRevA.84.053846,PhysRevA.91.043801}. In such an arrangement, a microwave quantum circuit could be constructed in which a YIG ferrimagnet is deposited on top of a superconducting coplanar waveguide resonator \cite{PhysRevLett.111.127003,PhysRevLett.123.107702,PhysRevLett.123.107701}. The resonant frequencies of the ferromagnet and the superconducting resonator are several GHz, and the coupling strength between the magnon and the microwave photon ranges from tens of megahertz to hundreds of megahertz. The decay rates of the resonator and the YIG ferromagnet can be as low as several megahertz, which are within reach of state-of-the-art experimental technologies. We mention that the microwave parametric driving required for implementing our scheme could also be realized by inserting a superconducting quantum interference device (SQUID) into the  superconducting resonator, which has been widely used in experiments \cite{PhysRevLett.119.023602,PhysRevX.7.041011,zhong2013squeezing}. Regarding the mechanical part, a microwave optomechanical subsystem needs to be further constructed and integrated into the above setup, where the superconducting microwave resonator is coupled to a mechanical oscillator \cite{wollman2015quantum,toth2017dissipative,PhysRevLett.123.183603} that is electrically pumped by a periodic drive for activating the MPA \cite{1438421,LUO2006139}. Typically experimental values of the mechanical frequency is tens of megahertz, and a mechanical quality factor exceeding $10^8$ has been demonstrated recently in nanomechanical resonators \cite{PhysRevLett.116.147202,tsaturyan2017ultracoherent,ghadimi2018elastic}. For the detection of the entanglement, measurement of the corresponding CMs would be performed on the hybrid system, where homodyne detection on the cavity output is required as discussed in Ref.~\cite{PhysRevLett.121.203601}.

Although the dual parametric amplification has been used to enhance the bipartite entanglement in an optomechanical setting \cite{yang2024multi}, extending it to a tripartite quantum system would undoubtedly increase the difficulty and complexity for manipulating the entanglement. Specifically, it might be more complicated to find an optimized parameter regime where the tripartite entanglement can be maximized in the simultaneous presence of both parametric drivings. Meanwhile, beyond the classes of bipartite entanglement, quantum entanglement shared by more than two partitions plays a more essential role in the development of programmable quantum networks and controllable dense coding \cite{armstrong2012programmable,cai2017multimode,PhysRevLett.90.167903}. Therefore, exploring effective methods to achieve strong and robust enough multipartite entanglement is of practical significance. Our work proposes for the first time to employ the synergistic strategy on basis of the dual parametric amplification to enhance and manipulate the tripartite entanglement. The application of the dual parametric amplification could also bring more degrees of freedom for the manipulation of entanglement, as we demonstrated in Fig.~\ref{fig5}. Furthermore, as a general recipe, the generalization of our method may also inspire the use of dual parametric amplification to manipulate various quantum effects in multiple fields of physics.

In conclusion, we proposed an effective method for manipulating and enhancing entanglement in a hybrid cavity-magnon optomechanical system. The scheme relies on applying an OPA or MPA modulation that provides enhanced coupling required for improving entanglement, based on which we achieved a considerable enhancement in the tripartite entanglement. In particular, we showed that the simultaneous application of phase-matched OPA and MPA enables further enhancement of the entanglement on top of that in the case of individually applying OPA or MPA. The underlying physics is verified and attributed to the constructive interference of the OPA and MPA when their phases are suitably matched. We envision our results to provide a flexible way for manipulating multipartite entanglement as well as other quantum effects in the hybrid system, and to pave the way for the enhancement and protection of fragile quantum resources.

\section{Acknowledgements}
The authors thank Dr. Dong-Yang Wang for valuable discussions. H.J. is supported by the National Natural Science Foundation of China (NSFC) (Grant Nos. 11935006 and 12421005), HNQSTIT project  (Grant No. 2022112) and the Henan Science and Technology Major Project of the Department of Science and Technology of Henan Province (Grant No. 241100210400). L.M.K. is supported by the NSFC (Grant Nos. 12247105, 12175060, 12421005 and 11935006) and HNQSTIT project (Grant No. 2022112). 
L.Y.J. is supported by the Joint Fund of Henan Province Science and Technology R$\&$D Program (Grant No. 225200810071) and the Science and Technology Major Project of Henan Province (Grant No. 231100220800).
Y.W. is supported by the NSFC (Grant No. 12205256), the Henan Provincial Science and Technology Research Project (Grant No. 232102221001) and the Doctoral Research Foundation (Grant No. 2022BSJJZK18). 
J.L.W. is supported by the NSFC (Grant No. 12304407) and the China Postdoctoral Science Foundation (Grant Nos. 2023TQ0310 and GZC20232446). Y.F.J. is supported by the NSFC (Grant No. 12405029). T.X.L. is supported by the NSFC (Grant No. 12205054).

\appendix
\section{Long-time dynamics of the cavity-field amplitude}\label{appenA}
In the main text, we have derived the analytical approximation for the steady-state cavity-field mean value $\langle\hat{a}(t)\rangle$ in the presence of OPA. On that basis, we plot in Fig.~\ref{fig4}(a) the long-time dynamics of the cavity-field amplitude $|\langle a(t)\rangle|$ for different parametric phases $\theta_{c}=0$, $\pi/2$ and $\pi$. As seen, $|\langle a(t)\rangle|$ is periodically modulated by the OPA, and the maximal value of $|\langle a(t)\rangle|$ remains unchanged with varying $\theta_{c}$. In addition to the analytical method, we can also examine the system dynamics by numerically solving the nonlinear differential equations (\ref{ssmve}). Figure~\ref{fig4}(b) shows the numerical result of the time evolution of $|\langle a(t)\rangle|$ under OPA, which is in accordance with the analytical result in Fig.~\ref{fig4}(a). Note that the deviation between the analytical approximation and the exact numerical result may be due to the fact that more higher-order terms should be included in Eq.~(\ref{eqA1}) for a better fitting. Nonetheless, a first-order approximation in Eq.~(\ref{eqA1}) is enough to provide us with a proof-of-principle study of the nonlinear dynamics. For the case of individually applying MPA, $|\langle a(t)\rangle|$ can also be periodically modulated (not shown here), and the maximal $|\langle a(t)\rangle|$ remains steady with varying $\theta_{m}$, which is due to the fact that the mechanical driving phase only affects the oscillation frequency of the cavity field. In contrast, in the simultaneous presence of both OPA and MPA, as shown in Fig.~\ref{fig4}(c), $|\langle a(t)\rangle|$ is modulated by the phase difference $\Delta_{\theta}$, exhibiting a typical interference behavior. Particularly, by comparing Figs.~\ref{fig4}(b) and \ref{fig4}(c) we see that the maximal value of $|\langle a(t)\rangle|$ for $\Delta_{\theta}=0$ is higher than that of individually applying the optical parametric driving, which is a consequence of the synergistic enhancement of the optomechanical coupling by both parametric drivings with matched phases (satisfying $\Delta_{\theta}=0$). In addition, we also examine the dynamics when the parametric drivings are absent, which display extremely slight oscillations (see gray dashed curves in Figs.~\ref{fig4}(b) and \ref{fig4}(c)). This implies that the parametric drivings employed in our scheme are prerequisite for modulating and enhancing the entanglement.

\bibliography{sample1.bib}

\end{document}